\documentclass[aps,pra,groupedaddress,superscriptaddress,twocolumn,showpacs]{
revtex4-1}

\usepackage[utf8x]{inputenc}
\usepackage{color}
\usepackage{bbm} 

\usepackage{amsfonts,amsmath,amssymb,stmaryrd}

\usepackage{graphicx}

\usepackage{subfigure}  

\usepackage{bbm} 

\usepackage{hyperref}

\usepackage{bbm}

\usepackage{epsfig}

\usepackage{mathrsfs}

\usepackage{verbatim}

\usepackage{ulem}	

\renewcommand{\l}{\left(}
\renewcommand{\r}{\right)}
\newcommand{\bra}[1]{\langle#1|}
\newcommand{\ket}[1]{|#1\rangle}

\renewcommand{\H}{\hat{\mathcal{H}}}
\renewcommand{\c}{\hat{c}}

\newcommand{\cd}{\hat{c}^\dagger}

\newcommand{\rh}{\hat{\rho}}

\newcommand{\rrh}{\uline{\rho}}

\newcommand{\LL}{\uuline{\mathcal{L}}}

\newcommand{\LLCP}{\uuline{\rm LCP}}

\newcommand{\hc}{\text{h.c.}}

\usepackage{array}

\usepackage{bm}	
\renewcommand{\vec}[1]{\bm{#1}}

\begin{document}
\normalem	

\title{Topological order of mixed states in correlated quantum many-body systems}

\author{F. Grusdt}
\affiliation{Department of Physics, Harvard University, Cambridge, Massachusetts 02138, USA}

\begin{abstract}
Topological order has become a new paradigm to distinguish ground states of interacting many-body systems without conventional long-range order. Here we discuss possible extensions of this concept to density matrices describing statistical ensembles. For a large class of quasi-thermal states, which can be realized as thermal states of some quasi-local Hamiltonian, we generalize earlier definitions of density matrix topology to generic many-body systems with strong correlations. We point out that the robustness of topological order, defined as a pattern of long-range entanglement, depends crucially on the perturbations under consideration. While it is intrinsically protected against local perturbations of arbitrary strength in an infinite closed quantum system, purely local perturbations can destroy topological order in open systems coupled to baths if the coupling is sufficiently strong. We discuss our classification scheme using the finite-temperature quantum Hall states and point out that the classical Hall effect can be understood as a finite temperature topological phase. 
\end{abstract}

\date{\today}

\maketitle

\section{Introduction}
Thermodynamics represents one of the most powerful theories, allowing to use simple microscopic models for the description of far more complex situations in reality. It relies on the existence of universality, i.e. the fact that entire classes of microscopic models give rise to the same macroscopic physical behavior. In a hydrodynamic regime these macroscopic properties can be described by a few order parameters. Therefore two microscopic states belong to the same universality class only if they have the same order parameters. Based on this idea, Ginzburg and Landau put forward a powerful theory of universality \cite{Ginzburg1950} and showed that ordering of a system is associated with the spontaneous breaking of a corresponding symmetry. The resulting universality classes can be classified by their symmetry properties alone. 

The quantum Hall effect \cite{Vonklitzing1980} provides an example for a quantum state of matter which does not break any symmetry at zero temperature. Nevertheless it can be distinguished from other symmetric states in this regime by its universal transport properties, hinting at an underlying ordering principle. Indeed it was realized that the global structure of the wavefunction in the quantum Hall effect has non-trivial topology \cite{Laughlin1981,Thouless1982,KOHMOTO1985}. This defines a new set of universality classes going beyond Ginzburg's and Landau's paradigm \cite{Wen2004}.

\begin{figure}[b!]
\centering
\epsfig{file=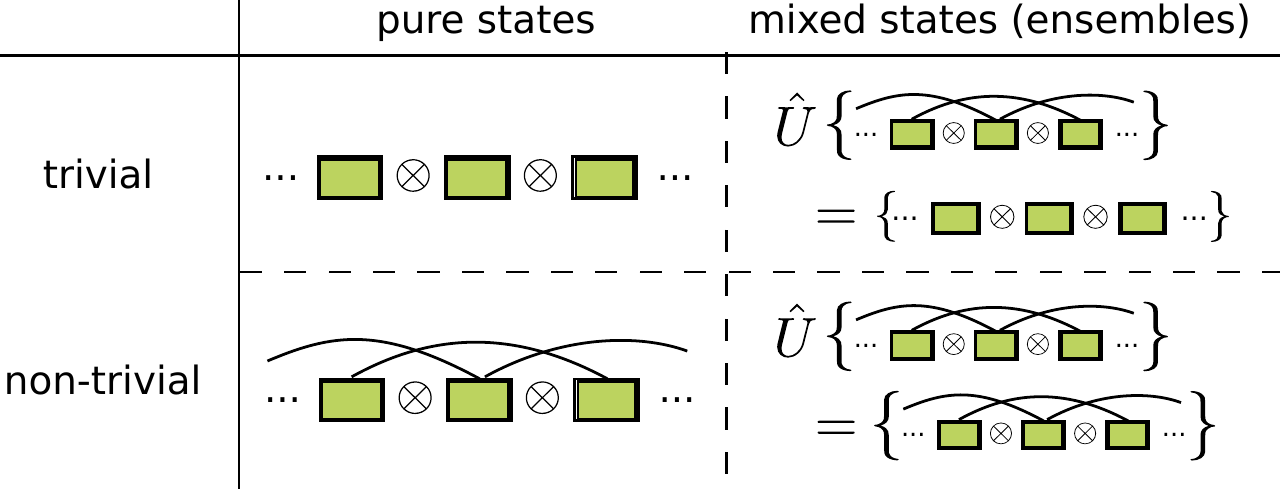, width=0.45\textwidth}
\caption{In a pure quantum state topological order corresponds to a pattern of long-range entanglement. For an ensemble of quantum states, all from a certain subspace of the full Hilbertspace, the notion of topological order can be generalized by allowing arbitrary basis changes $\hat{U}$ \emph{within} this restricted subspace. If this is sufficient to map the entire ensemble on an ensemble of topologically trivial states, the ensemble can be considered topologically trivial.}
\label{DMtopOrder}
\end{figure}

The ground states of closed quantum systems can be described by pure states. Interestingly, even if two wavefunctions are indistinguishable by investigating all their local order parameters, their global topology may not be the same. For pure states without long-range correlations in closed quantum systems, a unified theory \cite{Chen2010} of such topological order \cite{Wen1995} has been developed during the past decades \cite{Laughlin1981,Thouless1982,KOHMOTO1985,Laughlin1983,Wen1995,Kitaev2006,Ryu2010,Chen2010}. The key ingredient is the pattern of the non-local entanglement in a wavefunction, see Fig.~\ref{DMtopOrder}, which is completely robust to local perturbations of arbitrary (but finite) strength in an infinitely large system \cite{Chen2010}. In essence this means that two states are equivalent if they can be transformed into one another by arbitrary local basis changes. Because such basis changes are described by local unitary (LU) operations we refer to the resulting scheme as the LU classification \cite{Chen2010}.

Mixed states, on the other hand, can be understood as a statistical ensemble of pure states, and in principle each of these can have long-range entanglement -- i.e. topological order. Therefore a much richer topological structure should be expected for density matrices than for pure states. Currently, however, only little is known about this structure and the case of correlated many-body systems is widely unexplored. One of the main challenges is that, in general, the statistical ensembles under consideration contain pure states which cannot be written as ground states of a gapped local Hamiltonian and to which the LU scheme thus cannot be straightforwardly applied \cite{Chen2010}.  

For Gaussian states of fermions an exhaustive classification scheme has been suggested \cite{Diehl2011,Bardyn2013}. It can be understood by writing Gaussian density matrices as thermal states of free Hamiltonians. The latter have been grouped into a set of topologically inequivalent universality classes \cite{Ryu2010}. This reduction to non-interacting particles effectively solves the problem of excited states with long-range correlations, because all topological properties are derived directly from the ground state of the free Hamiltonians. Other approaches for understanding topological order in open quantum systems rely on geometric phases and generalizations thereof \cite{Uhlmann1986,Avron2011,Avron2012,Rivas2013a,Huang2014,Nieuwenburg2014,Viyuela2014,Budich2015,Budich2015a,Albert2016}, or on the closely related concept of macroscopic (many-body) polarization \cite{Linzner2016}.

In this paper we introduce a generalization of the LU classification scheme to density matrices describing correlated many-body systems. First we consider a large class of quasi-thermal states $\rh$, for which $\hat{h} =- \log \rh$ defines a (quasi) local Hamiltonian. This allows a physical interpretation of such states as true thermal states of $\hat{h}$. (As usual, $\hat{h}$ is called local, if there exists a finite length scale beyond which no couplings are allowed; It is called-quasi local, if all couplings decay at least exponentially with distance beyond a similar length scale.) Then we generalize the LU classification scheme for arbitrary density matrices. Here the essence is to identify patterns of the long-range entanglement in the statistical ensemble, which is invariant under continuous changes of $\rh$. As illustrated in Fig.~\ref{DMtopOrder}, a density matrix is called topologically trivial if a local basis exists in which, up to adiabatic variations, it describes a statistical ensemble of product states (we will formalize below what this means more precisely). 

Aside from the theoretical interest in a classification scheme of topological order of arbitrary mixed states, we are motivated by the challenge of simulating physical systems with non-trivial topology numerically. Understanding which patterns of the non-local entanglement can exist is important for choosing suitable variational states \cite{Vidal2004,Verstraete2004,Eisert2010RMP} which can capture the topological properties of the density matrix. 

In closed quantum systems, topological order defined as a pattern of the long-range entanglement \cite{Wen1995} is robust to local manipulations of the state \cite{Chen2010}. In this paper we use the generalized LU classification scheme to study how density matrices are affected by such local perturbations. We argue that one has to carefully distinguish between mixed states in closed quantum systems and in more general driven-dissipative systems coupled to baths. While robust intrinsic topological order (which does not require any symmetries) can exist in the former case, we argue that its robustness is absent in the latter case, although weak local perturbations have no effect. 

The LU classification scheme for density-matrix topological order introduced here, is a direct generalization of the scheme by Diehl et al. \cite{Diehl2011,Bardyn2013}. In contrast to the earlier approaches, our description is not limited to Gaussian density matrices but can be applied to arbitrary correlated many-body states. It is also related to the quantum circuit approach put forward by Hastings \cite{Hastings2011}, which represents an alternative way of classifying the long-range entanglement in a density matrix.

The paper is organized as follows. In Sec.~\ref{sec:ReviewPureStates} a brief review of the LU classification scheme for pure states is given. Sec.~\ref{sec:ClosedSystems} is devoted to a discussion of thermal states in closed quantum systems. In Sec.~\ref{sec:LUtopOrderDM} we introduce our main result and generalize the LU scheme to more general density matrices. Two concrete examples of density matrices with topological order, describing the Hall effect, are also presented. Open quantum systems coupled to Markovian baths are discussed in Sec.~\ref{sec:OpenSystems}. We close with a summary and by giving an outlook in Sec.~\ref{sec:outlook}.

\section{Brief review of topological order of pure states}
\label{sec:ReviewPureStates}

\subsection{Geometric phases}
\label{sec:Geo}
Arguably, the integer quantum Hall system constitutes the most famous topological phase of matter. Its intriguing (adiabatic) transport properties can be directly related to a geometric quantity defined by the Berry curvature \cite{Berry1984,Zak1989,Xiao2010} for the Bloch wavefunctions $u_\alpha(\vec{k})$ of the bands $\alpha$ occupied by electrons,
\begin{equation}
\mathcal{F} =  \sum_\alpha \epsilon_{\mu \nu} \partial_{\mu} \bra{u_\alpha(\vec{k})} \partial_{\nu} \ket{u_\alpha(\vec{k})}.
\label{eq:BerryCurvature}
\end{equation}
The integral of the Berry curvature over the Brillouin zone (BZ), defining the Chern number $\mathcal{C}$, represents an integer-quantized topological invariant \cite{Thouless1982},
\begin{equation}
\mathcal{C} = \int_{\rm BZ} d^2k ~ \mathcal{F} \in \mathbb{Z}.
\label{eq:ChernNumber}
\end{equation}

The Chern number \eqref{eq:ChernNumber} allows to distinguish different many-body states, in this case Slater determinant wavefunctions defined from single-particle orbitals of a set of occupied bands. Therefore the topological order \cite{Wen1995} described by the Chern number provides a universality principle: microscopic many-body states which have the same Chern number belong to the same topological universality class. Two comments are in order, however. Firstly, only states with short-range correlations are classified in this way, because the definition of the Chern number requires a gapped state to begin with \footnote{If there is no gap, a different gauge choice can be used to obtain a different Chern number, making the latter ill-defined.}. Secondly it should be noted that additional topological quantum numbers may exist which allow to distinguish further between different states with the same Chern number.

\subsection{Long-range entanglement}
\label{sec:LRE}
More recently a refined theory of topological order has been developed, which applies more generally and -- unlike the Chern number -- is no longer based on geometry \cite{Chen2010}. Instead, it relies on the quantum mechanical entanglement of the many-body wavefunction. Two gapped states described by wavefunctions $\ket{\psi_1}$ and $\ket{\psi_2}$ in an infinitely large system are topologically equivalent iff they can be transformed into one another by a finite time evolution with a local Hamiltonian, a so-called LU transformation \cite{Chen2010}:
\begin{equation}
\hat{U}_L = \mathcal{T} \exp \l  - i \int_0^1 d\tau ~ \tilde{\mathcal{H}}(\tau) \r.
\label{eq:defLU}
\end{equation}
I.e. $\ket{\psi_1} \simeq \ket{\psi_2}$ iff $\ket{\psi_1} = \hat{U}_L \ket{\psi_2}$ for some local $\tilde{\mathcal{H}}(\tau)$. Here a state is called gapped, if it can be written as the ground state of a gapped local Hamiltonian; An operator is called local if it can be written as sum of operators $\hat{h}_j$ which are bounded and act on a local Hilbertspace, 
\begin{equation}
\tilde{\mathcal{H}} = \sum_j \hat{h}_j.
\end{equation}
Non-local terms with a coupling strength decaying exponentially with distance are acceptable, and in this case the Hamiltonian $\tilde{\mathcal{H}}$ is called quasi-local.

The effect of LU transformations can be understood as a local change of the basis. Therefore the LU scheme distinguishes only between wavefunctions with different non-local properties. States with conventional long-range order of a local order parameter are not distinguished however, because arbitrary local basis changes can easily destroy such long-range order.

The topologically trivial class is defined by the set of states which can be related to a product state by a LU transformation. I.e. $\ket{\psi_0}$ is trivial iff for some $\hat{U}_L$
\begin{equation}
\hat{U}_L  \ket{\psi_0} = \bigotimes_j \ket{\phi_j},
\end{equation}
where $\ket{\phi_j}$ are states in a local Hilbertspace $H_j$ labeled by $j$. Non-trivial states, on the other hand, cannot be written as product states and this fact can be reflected for example in their topological entanglement entropy \cite{Kitaev2006}. This clarifies why topological order represents a pattern of the long-range entanglement in a quantum state \cite{Wen2004}.

A comment is in order why the LU scheme can only classify gapped states. Ground states of gapped Hamiltonians fulfill an area law for their entanglement \cite{Hastings2007}. As shown by Kitaev and Preskill \cite{Kitaev2006}, this is essential to define the sub-leading correction to the entanglement entropy which stems from the non-local topological order. For low-energy states in a gapless Hamiltonian, on the other hand, the entanglement entropy can have sub-leading corrections which scale logarithmically with the volume, see Ref.~\cite{Eisert2010RMP}. This makes it difficult to distinguish between intrinsic topological entanglement and contributions due to quantum fluctuations delocalized over the system. 

The Chern number \eqref{eq:ChernNumber} is only one example of a topological quantum number, which is invariant under LU transformations and thus allows to easily distinguish states from different universality classes. A direct relation between the LU classification discussed above and the Chern number classification of topological states was established in in Ref.~\cite{Brouder2007}. The authors of \cite{Brouder2007} have shown that Wannier functions can be exponentially localized (i.e. the Slater determinant state of a band insulator can be written as a product state) if and only if the Chern number vanishes. The topological entanglement entropy \cite{Kitaev2006} is a second example for a topological invariant. Note however that it is not equivalent to the Chern number: For example, integer quantum Hall states have no topological entanglement entropy but a non-vanishing Chern number.

From the definition of the LU scheme in Eq.\eqref{eq:defLU} it follows that topological order is robust against local perturbations of arbitrary strength in an infinite system, \emph{not} limited by the energy gap above the ground state. For sufficiently large but finite systems, the perturbations need to be finite and small compared to the system size. In this case the topological order is only robust for a finite time which scales like the system size. In this paper we will always consider the ideal limit of infinitely large systems, however. The robustness of topological order will be illustrated using a simple toy model below. 

Remarkably, the LU classification scheme explicitly includes the possibility to study topological order far-from equilibrium, because any local Hamiltonian $\tilde{\mathcal{H}}(\tau)$ can be considered in the LU time-evolution. This property distinguishes the LU scheme from other approaches to define topological order, based for example on geometric phases in adiabatic evolutions of a quantum system \cite{Thouless1982}. 

On the downside, the relation between topological order, defined in a rather abstract way as a pattern of the long-range entanglement \cite{Wen1995,Chen2010}, and directly observable experimental consequences becomes more involved. While geometric phases are directly related to an adiabatic response of the system, giving rise for example to the strictly quantized Hall current, the long-range entanglement itself is challenging to detect. Note that the robustness of topological order and the insensitivity of local observables to the latter goes hand in hand. However, new detection schemes which are sensitive to quantum-mechanical entanglement have recently been developed \cite{Abanin2012Ent,Daley2012,Islam2015,Pichler2016}, and we expect that this will provide new ways of directly detecting topological order in the future.

\subsection{Robustness of topological order: Toy model}
\label{sec:toyModel}
\begin{figure}[t!]
\centering
\epsfig{file=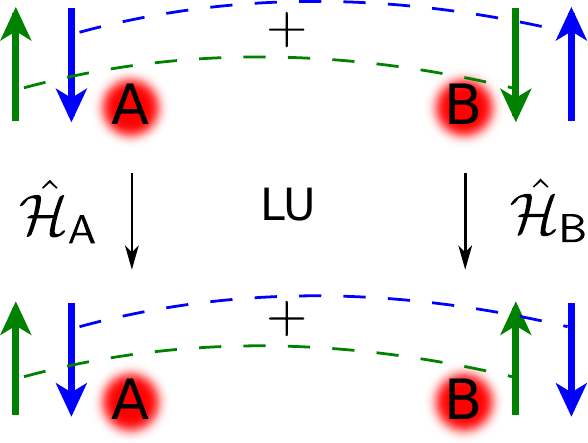, width=0.22\textwidth} $\qquad$
\caption{Illustration of topological order representing non-local (long-range) entanglement in a wavefunction, shared by two spatially separated parties (A and B) in this case. The long-range entanglement (i.e. the topological order) is robust to arbitrary local basis changes, corresponding to LU transformations generated by local Hamiltonians $\H_{\mathsf A,B}$.}
\label{LU_aliceBob}
\end{figure}

To explain the robustness of topological order in a closed quantum system in the most fundamental way, we use the toy model shown in Fig.~\ref{LU_aliceBob}. We consider the situation envisioned by Einstein, Podolsky and Rosen \cite{Einstein1935}, where two spatially separated parties A and B share a Bell state, e.g.
\begin{equation}
\ket{\Psi^+} = \frac{1}{\sqrt{2}} \Bigl( \ket{ \! \uparrow}_A \ket{ \! \downarrow}_B + \ket{ \! \downarrow}_A \ket{ \! \uparrow}_B \Bigr).
\end{equation}
The non-local entanglement can be characterized by the entanglement entropy of either of the subsystems, $S_A=-\text{tr} \rh_A \log \rh_A = \log 2$.

In the toy model, local unitary perturbations are described by Hamiltonians $\H_{A,B}$ acting separately on the two subsystems $A$ and $B$. They give rise to the following LU transformations 
\begin{equation}
\hat{U}_{L}^{n} = \mathcal{T} \exp \l - i \int_0^1 d\tau ~ \H_n(\tau) \r, \qquad n=A,B.
\label{eq:LUtransfAB}
\end{equation}
By redefining the local bases, $\ket{\tilde{\uparrow}}_n = \hat{U}_{L}^{n} \ket{ \! \uparrow}_n$ and $\ket{\tilde{\downarrow}}_n = \hat{U}_{L}^{n} \ket{ \! \downarrow}_n$, it is easy to see that the non-local entanglement entropy $\tilde{S}_A=\log 2$ of $\ket{\tilde{\psi_0}} = \hat{U}_{L}^{n} \ket{\Psi^+}$ is conserved under LU transformations \eqref{eq:LUtransfAB}. 

In a many-body system with topological order, as considered in Ref.~\cite{Chen2010}, the situation is very similar. In this case the entanglement entropy $S=\alpha L - \gamma + \mathcal{O}(1/L)$ between two regions separated in space has a constant topological contribution $-\gamma$ and an extensive contribution $\alpha L$ proportional to the area $L$ of the cut separating the two subsystems \cite{Kitaev2006}. In our toy model, we have introduced two decoupled subsystems which we can think of as being in two separate locations. We assume that they are sufficiently far apart from one another, such that there are no couplings across a boundary separating the two parts. This corresponds to the case $\alpha=0$, and all entanglement entropy is non-local in this sense.

\section{Topological order of thermal states in closed systems}
\label{sec:ClosedSystems}
We start by discussing topological order of thermal states in a closed quantum system. The analysis in this section provides the basis for understanding the meaning of topological order for more general mixed states. The density matrix of a thermal state can be written as
\begin{equation}
\rh_T = Z^{-1} e^{- \beta \hat{\mathcal{H}}},
\label{eq:ThermalState}
\end{equation}
where $\hat{\mathcal{H}}$ is a local Hamiltonian, $Z={\rm tr}~ e^{- \beta \H}$ and $\beta = 1/k_B T$ where $T$ is the temperature. In fact, we will use Eq.~\eqref{eq:ThermalState} as our definition of when a state is called thermal. 

In this section we assume throughout that the system is closed, i.e. the dynamics of the thermal state can be described by some local (possibly time-dependent) Hamiltonian $\tilde{\mathcal{H}}(\tau)$; no couplings to external baths are allowed, which would lead to non-unitary dynamics of the systems. Note that we do not make any assumptions about how the thermal state \eqref{eq:ThermalState} was initially prepared, which in practice may require couplings to external baths. We only make assumptions about the time-evolution afterwards, when the system is closed.

To define topological equivalence classes of thermal density matrices in the spirit of the LU scheme, we have to identify a set of manipulations which leaves their global structure invariant. A natural choice is to consider local basis changes, which can be described by a time-evolution with a local Hamiltonian, see Eq.\eqref{eq:defLU}.

In the following we consider a closed quantum system initially prepared in a thermal state $\rh_T^{(0)}$. Because there is no coupling to external baths, the subsequent dynamics can be described by a unitary time evolution, governed by a local Hamiltonian $\tilde{\mathcal{H}}(\tau)$. This corresponds to the action of a LU transformation on the density matrix, 
\begin{equation}
\rh_T(t) = \hat{U}_L^\dagger(t) \rh_T^{(0)} \hat{U}_L(t),
\label{eq:actionLUrhoT}
\end{equation}
where $\hat{U}_L^\dagger(t)=\mathcal{T} \exp [- i \int_0^t \tilde{\mathcal{H}}(\tau)]$. To compare different density matrices, and ask whether they have the same topological order, we will distinguish between two scenarios now. 

In the first case (globally thermal state) we consider a situation where the \emph{entire} quantum system is initially described by the thermal state $\rh_T^{(0)}$, i.e. not only local but also global observables can be calculated using Eq.\eqref{eq:ThermalState}. This situation is expected, for example, when a system is initially coupled to a large reservoir with which it thermalizes. In this case we can compare different density matrices describing the entire system.

In the second case (locally thermal state) we assume that only \emph{local} observables can be described by the thermal state \eqref{eq:ThermalState}, i.e. the reduced density matrices of local subsystems are thermal. In this case we will restrict our analysis to the comparison of different reduced density matrices of the same local subsystem. Note however that the size of the subsystems $\ell$ can be considered to be much larger than the correlation length $\xi$, while still being much smaller than the system size $L$,
\begin{equation}
\xi \ll \ell \ll L \to \infty.
\label{eq:rangeEll}
\end{equation}
This is the limit we will consider from now on.

\subsection{Globally thermal states}
\label{sec:global}
By definition, the LU transformations $\hat{U}_L(t)$ cannot change the structure of the long-range entanglement in the globally thermal state $\rh_T^{(0)}$. Therefore the topological order of $\rh_T(t)$ is the same for all times $t$. Physically this statement can be understood from the fact that it takes an infinite amount of time until long-range entanglement can build up across the entire, infinitely large system. This is a manifestation of the Lieb-Robinson bound for the spreading of entanglement in the presence of purely local couplings \cite{Bravyi2006}.

A natural definition of topological order for globally thermal quantum states can thus be given by applying the LU classification scheme separatly to every state in the ensemble:\\\emph{\textbf{Definition (global topological order):} Two globally thermal states $\rh_{T}^{(1)}$ and $\rh_{T}^{(2)}$ in a closed quantum system are topologically equivalent $\rh_{T}^{(1)} \sim \rh_{T}^{(2)}$ iff they are related by LU transformations,}
\begin{equation}
\rh_{T}^{(2)} = \hat{U}^\dagger_L \rh_{T}^{(1)} \hat{U}_L.
\label{eq:LUdensityMatrices}
\end{equation}
Because LU transformations define a mapping between different thermal states, and since the inverse $\hat{U}_T^\dagger$ also defines a LU transformation, one easily confirms that Eq.\eqref{eq:LUdensityMatrices} defines an equivalence relation. For pure states, the classification is equivalent to the LU scheme.

However, because unitary transformations leave the spectrum of the operator $\rh_T$ invariant, the definition in Eq.\eqref{eq:LUdensityMatrices} only allows to compare density matrices with a fixed global spectrum. While this confirms our intuition that the topological order of an ensemble of quantum states in a closed system can be much richer than for pure states, it makes the definition of little practical use. 

Before proceeding to a less restrictive definition of topological order for locally thermal states, we would like to emphasize the strength of the definition above. In direct analogy to the robustness of topological order in pure states derived from the LU classification, see also Sec.~\ref{sec:toyModel}, we can make a similar statement for globally thermal density matrices: \emph{Global topological order of thermal states in a closed quantum system is robust to local perturbations of arbitrary strength in an infinite system}. We will discuss in Sec.~\ref{sec:OpenSystems} that the assumption of a closed quantum system is crucial for this result to hold.

\subsection{Locally thermal states}
\label{sec:local}
The time-evolution in Eq.\eqref{eq:actionLUrhoT} cannot change the long-range entanglement globally. Nevertheless, the structure of the entanglement in a reduced density matrix of a local subsystem can change completely after a finite time related to the size $\ell$ of the subsystem. In other words, the topological properties of the system may change faster on shorter length scales than on longer ones. 

To introduce a precise notion of topological order in a subsystem, we will now define adiabatic deformations of a density matrix which leads us to a more restrictive definition of topological order than given above for globally thermal states. As an important example we discuss adiabatic time-evolutions of non-integrable (i.e. thermalizing) quantum systems, and argue that these lead to adiabatic deformations of the reduced density matrix of a local subsystem.

\subsubsection{Adiabatic changes of thermal density matrices}

\begin{figure}[t!]
\centering
\epsfig{file=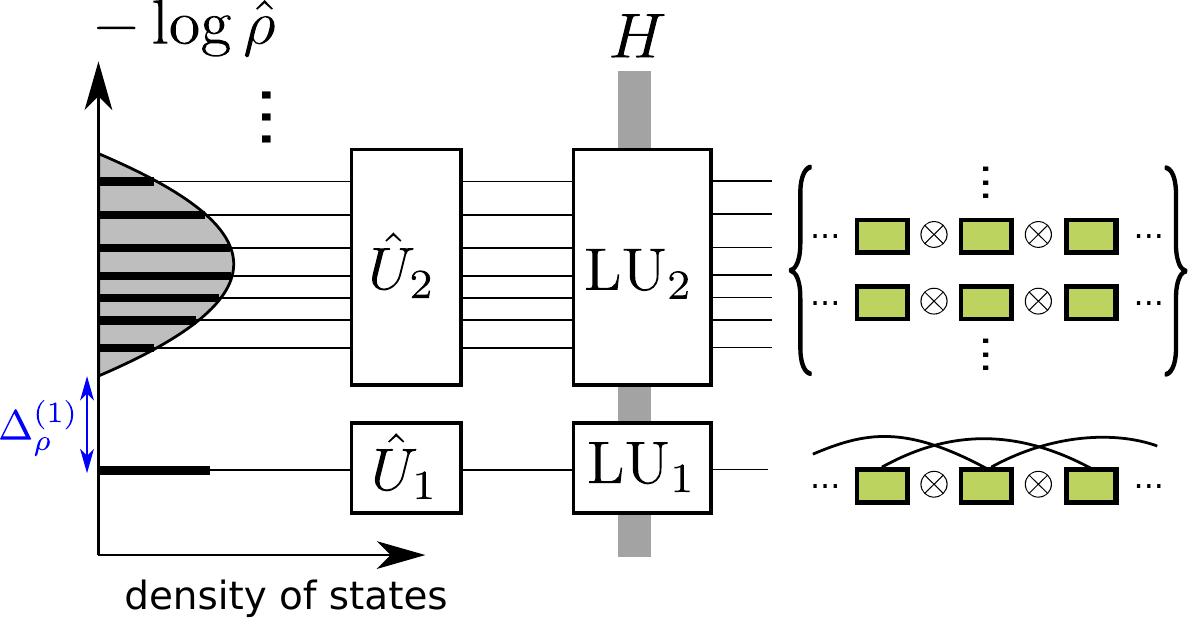, width=0.45\textwidth}
\caption{The spectrum of $- \log \rh$ corresponding to a generic density matrix $\rh$ can be divided into manifolds $\mu$ of states separated by gaps $\Delta_\rho^{(\mu)}$. Adiabatic deformations of the density matrix $\rh$ leave this spectral structure invariant. Two density matrices with the same spectral structure are defined to be topologically equivalent if all manifolds are equivalent; Two manifolds of states from different density matrices are topologically equivalent if they can be transformed into one another by a combination of arbitrary unitary transformations $\hat{U}$ within the manifold, and a $\rm LU$ transformation acting on the entire Hilbert space $H$. We sketch an example where the first manifold is long-range entangled and the second manifold is equivalent to an ensemble of topologically trivial states.}
\label{DMdefTopOrd}
\end{figure}

To make a definition of adiabatic changes (or continuous deformations) of a locally or globally thermal density matrix $\rh_T$, let us consider the spectrum of the operator $- \log \rh_T = \beta \H$ in a generic many-body system. In general it consists of $\mu=1...M$ manifolds of states, separated by gaps $\Delta_\rho^{(\mu)}$ in the spectrum, see Fig.~\ref{DMdefTopOrd}. The number of states in each manifold, and the number of manifolds separated by gaps define the \emph{spectral structure} of the density matrix. Note that the choice of calculating the spectrum of $- \log \rh_T$ is in principle arbitrary, because the spectral structure of other operators like $\rh$ or $\rh^2$ is equivalent. It is motivated by the fact that $- \log \rh_T$ yields the Hamiltonian (up to the factor $\beta$) in the case of thermal states, which connects our classification scheme directly to approaches developed for pure eigenstates of the Hamiltonian. 

When $- \log \rh_T$ actually describes a Hamiltonian, adiabatic changes of the latter have to be slow compared to the gaps $\Delta_\rho^{(\mu)}/\beta$. In this case there can be no transitions between different manifolds and the population within each manifold is conserved, i.e. the spectral structure of the density matrix is invariant. This physical consideration motivates the following definition: \emph{A deformation of a density matrix $\rh \to \rh'$ is called adiabatic, if it leaves the spectral structure of the density matrix invariant.} Note that within the different  manifolds $\mu$ in the spectrum, the density matrix can change and non-trivial reorganizations may take place. 

As in the case of closed quantum systems, we assume that adiabatic changes of the state should not modify the topological order of a thermal state describing the local subsystem. This motivates the following definition:\\
\emph{\textbf{Definition (local topological order)}: Two locally thermal states $\rh_T^{(1)}$ and $\rh_T^{(2)}$ in a subsystem of a closed quantum system are topologically equivalent, $\rh_{T}^{(1)} \sim \rh_{T}^{(2)}$, iff they can be transformed into one another by adiabatic deformations. This requires $\rh_T^{(1,2)}$ to have the same spectral structure.}\\
Note that we restricted ourselves to reduced density matrices $\rh_T^{(1,2)}$ of local subsystems here, because as discussed in the previous section, in a closed quantum system the spectrum of the global density matrix cannot change. 

In the case when the local subsystem can be described by a pure quantum state, $\rh_T = \ket{\psi}\bra{\psi}$, the definition of local topological order coincides with the usual definition for pure states in a closed system. It is equivalent to the LU classification scheme \cite{Chen2010}, to which also the definition of global topological order simplifies in this limit. 

Unlike in the case of global topological order, local topological order is \emph{not} robust to arbitrary local perturbations in an infinite system. Because we consider only a local subsystem, strong perturbations can entangle it with the surrounding parts of the closed quantum system in a finite time. This leads to thermalization of the local subsystem and may cause changes of the spectral structure of the reduced density matrix.

On the other hand, \emph{local topological order remains robust to arbitrary local perturbations which are weak compared to the gaps $\Delta_\rho^{(\mu)}/\beta$}: They only lead to adiabatic changes of the reduced density matrix and hence leave its spectral structure invariant.

\subsubsection{Adiabatic evolutions of reduced density matrices in thermalizing systems}
As an important example which illustrates the relevance of adiabatic variations of density matrices, let us consider a generic closed quantum system which is non-integrable. Further we assume that the initial state is locally thermal. Therefore, after a global time evolution described by $\tilde{\mathcal{H}}(\tau)$ as in Eq.\eqref{eq:actionLUrhoT}, local observables are expected to thermalize after a finite time $t'>t$ \cite{Rigol2012}. Hence the reduced density matrix of a local subsystem, $\rh_T'$, is still given by a thermal state at time $t'$, but using the new Hamiltonian $\H' = \tilde{\mathcal{H}}(t)$ and a new temperature $\beta'$; $t'$ depends on the size of the subsystem $\ell$, which is assumed to be finite, see Eq.\eqref{eq:rangeEll}. 

To compare different thermal states describing the same local subsystem with different Hamiltonians $\H'$, we have to distinguish between two cases. When $\tilde{\mathcal{H}}(\tau)$ is varied sufficiently slowly (\emph{adiabatically}), the structure of the probability distribution described by the reduced density matrix $\rh_T'(t)$ does not change; i.e. the spectral structure of $\rh_T'(t)$ is left invariant. If, on the other hand, $\tilde{\mathcal{H}}(\tau)$ is varied too quickly (\emph{quench}), the structure of the reduced density matrix can change completely. In this case the local topological order, as defined above, can change.

\section{LU classification of density-\\matrix topological order}
\label{sec:LUtopOrderDM}
Now we introduce a more practical definition of local topological order in a density matrix. To this end we return to a discussion of the effect of adiabatic changes of the Hamiltonian $\tilde{\mathcal{H}}(\tau)$ on the reduced density matrix $\rh_T$ of a local subsystem. This leads us to a definition of density matrix topological order in terms of LU transformations, which we suggest to use more generally for a larger class of density matrices.

In this section we start by generalizing our discussion to quasi-thermal states. They will be defined as general density matrices $\rh$ with a local logarithm $\hat{h} = - \log \rh$ (exponentially decaying terms in $\hat{h}$ are acceptable, in that case $\hat{h}$ is quasi-local). Because such quasi-thermal density matrices can be interpreted as thermal states of the effective local Hamiltonian $\hat{h}$, the same considerations as in the case of true thermal states apply.  Note, however, that we can only think of $\rh$ as an effective thermal state as long as $\log \rh$ remains local; otherwise a meaningful distinction between local and non-local entanglement is impossible, which, however, is indispensable for defining topological order \cite{Wen1995}. Later we will generalize our definition of topological order and apply the LU classification scheme to arbitrary density matrices. In this section we do not discuss the robustness of density matrix topological order, so there is no need to distinguish between open and closed quantum systems. Rather, the goal is to identify the topological structure of a given density matrix in a physically meaningful way.

As discussed in the previous section, adiabatic changes of a quasi-thermal density matrix leave its spectral structure invariant. Within each manifold $\mu$, arbitrary dynamics can take place because of the absence of a gap. Their effect can be described by a unitary matrix $\hat{U}_\mu$ acting between states from the manifold $\mu$. In addition, the local Hamiltonian $\tilde{\mathcal{H}}(\tau)$ can mix quantum states from different manifolds, without changing their populations however. This is described by the action of a LU transformation. Here we assume that LU transformations act on a finite length scale $L_0 \geq \xi$ larger than the correlation length, but smaller than the size $\ell$ of the subsystem under consideration, $L_0 \ll \ell$. 

Therefore, as summarized in Fig.~\ref{DMdefTopOrd}, the effect of adiabatic variations of the quasi-thermal density matrix $\rh$ is a combination of the unitaries $\hat{U}_\mu$ acting within the manifolds, and LU transformations acting between them. These considerations lead us to the following definition of topological order in a quasi-thermal density matrix, which formalizes our definition of topological order in a locally thermal state given in Sec.~\ref{sec:local}.\\
\emph{\textbf{Definition (LU topological order)}: Two states $\rh_T^{(1)}$ and $\rh_T^{(2)}$ with the same spectral structure (manifolds $\mu=1,...,M$) are topologically equivalent, $\rh_{T}^{(1)} \sim \rh_{T}^{(2)}$, iff all their manifolds are topologically equivalent. A manifold of states $\Psi = \{ \ket{\psi_\mu^n} \}_{n=1...N_\mu}$ is topologically equivalent to $\Phi = \{ \ket{\phi_\mu^n} \}_{n=1...N_\mu}$, iff it can be transformed into $\Phi$ by a combination of a unitary transformation $U_\mu$ within the manifold, and a local unitary transformation $\hat{U}_L$ acting on the entire Hilbert space:}
\begin{equation}
\ket{\psi_\mu^n} = \hat{U}_L \sum_m  U_\mu^{n,m} \ket{\phi_\mu^m}.
\label{eq:defLUdm}
\end{equation}

One easily confirms that our definition of density matrix topology represents an equivalence relation: $\rh \sim \rh$ follows trivially by using the identity matrix, $U_\mu=1$, $\hat{U}_L=\hat{1}$; For $\rh_1 \sim \rh_2$ it follows that $\rh_2 \sim \rh_1$ by using $U_\mu'=U_\mu^\dagger$ and $\hat{U}_L'=\hat{U}_L^\dagger$; And from $\rh_1 \sim \rh_2$ (with $U_\mu$ and $\hat{U}_L$) and $\rh_2 \sim \rh_3$ (with $U_\mu'$ and $\hat{U}_L'$) it follows that $\rh_1 \sim \rh_3$ by using $U_\mu'' = U_\mu' U_\mu$ and $\hat{U}_L''=\hat{U}_L' \hat{U}_L$.

A density matrix is topologically trivial, iff all manifolds can be transformed into an ensemble of product states. This is illustrated for an ensemble from a single manifold in Fig.~\ref{DMtopOrder}. If at least one ensemble is topologically non-trivial, the entire density matrix contains a non-trivial structure of long-range entanglement, which cannot be eliminated by basis changes within the manifolds. 

The definition of LU topological order provided above can be applied more generally to density matrices which are not quasi-thermal (i.e. $\log \rh$ is non-local). In this case our arguments relying on adiabatic deformations of quasi-thermal states do not apply, and Eq.~\eqref{eq:defLUdm} provides a formal definition of density-matrix topological order.

Pure states correspond to density matrices with a particularly simple spectral structure: The pure state has quasienergy $- \log \rho = 0$ whereas for all other states $- \log \rho = \infty$. In this case $U_\mu=1$ and the density matrix LU scheme reduces to the original LU classification for pure states. Thermal states $\rh_T = e^{- \beta \H}/Z$ at finite temperatures, $0 < \beta < \infty$, all have the same topological classification, determined by the spectrum and eigenstates of $\H$. The state at $T=\infty$ is always topologically trivial.

A comment is in order about the nature of transitions between density matrices with different topology. This requires a change in the spectral structure, i.e. at least one of the spectral gaps $\Delta_\rho^{(\mu)} = 0$ has to close. Unlike in the case of pure ground states in a closed quantum system, a topological transition does not require the system to become critical. I.e. \emph{density matrix topological order does not classify different physical phases}. This phenomenology has been previously predicted for Gaussian systems by Diehl and co-workers \cite{Diehl2011,Bardyn2013} (see also discussion below). Nevertheless the pattern of long-range entanglement, i.e. the topological order in the density matrix, changes at the topological transition. 

We note that eigenstates of the density matrix for which the value of $- \log \rho$ is very large may only play a sub-dominant role in determining the properties of the ensemble. To take this effect into account, a refined definition of LU density-matrix topological order can be made where only the first few manifolds are considered.

\subsection{Example: topological order at finite $T$}
\label{sec:Example}
As a generic example we discuss the density matrix topological order of integer Chern insulators in two dimensions, i.e. the lattice versions of the integer quantum Hall effect \cite{HOFSTADTER1976,Haldane1988}. As in Sec.~\ref{sec:Geo} we consider fermions occupying Bloch bands $\ket{u_\alpha(\vec{k})}$ which are separated by a band gap $\Delta > 0$ from the unoccupied states $\ket{u_\beta(\vec{k})}$, first without interactions. We will discuss a generic class of topological transitions existing in these models at finite temperatures, when the ratio of the band gap to the band width is tuned. Along these lines we argue that the classical Hall effect can be understood as a manifestation of density-matrix topological order.

\subsubsection{Toy model: Two bands}
\begin{figure}[t!]
\centering
\epsfig{file=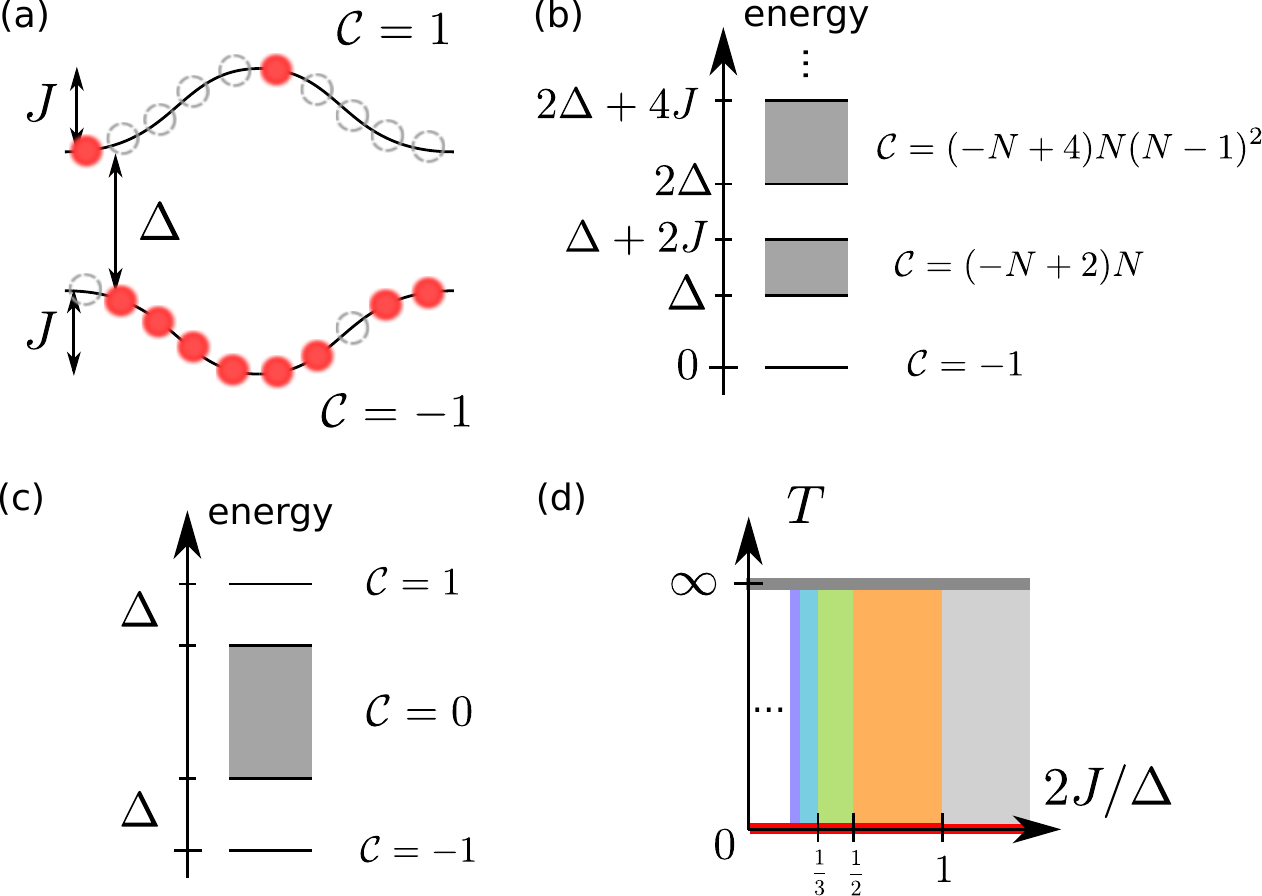, width=0.5\textwidth}
\caption{(a) As a toy model that possesses density-matrix topological order, we discuss a two-band integer Chern insulator ($N$ particles) at finite temperatures. (b) The spectral structure of the thermal density matrix, $- \log \rh = \beta \H$, has gaps $\Delta_\rho^{(\mu)}$ at low energies if the combined band width of the two bands is smaller than the band gap, $2 J < \Delta$. If on the other hand $2 J > \Delta$, only the first gap $\Delta_\rho^{1}$ remains open (c). The many-body Chern numbers $\mathcal{C}$ of the resulting manifolds of states are indicated. As a result we obtain the phase diagram shown in (d), where regions with different topological order are indicated by different colors.}
\label{ToyModelDMtopOrder}
\end{figure}

We start from the Haldane model \cite{Haldane1988}, which has two bands $\ket{u_\pm(\vec{k})}$ with energies $\epsilon_\pm(\vec{k})$ and opposite Chern numbers, $\mathcal{C}_+ = - \mathcal{C}_-$. To simplify our analysis, we assume that the energies of the two bands are related by $\epsilon_-(\vec{k}) = - \epsilon_+(\vec{k})$, and that they can be characterized by the band width $J$ and the band gap $\Delta$, see Fig.~\ref{ToyModelDMtopOrder} (a). We consider the case of half filling where the lowest band is completely filled at zero temperature, realizing a pure Chern insulator. To begin with we assume that the fermions are non-interacting.

Next we construct the spectrum of the many-body Hamiltonian
\begin{equation}
\H_0 = \sum_{\vec{k}} \sum_{\tau = \pm}  \epsilon_\tau(\vec{k}) ~ \cd_{\vec{k},\tau} \c_{\vec{k},\tau},
\end{equation}
which determines the spectral structure $- \log \rh = \beta \H_0$ of the thermal density matrix. As shown in Fig.~\ref{ToyModelDMtopOrder} (b), the first manifold contains only the ground state which is separated by a gap $\Delta_\rho^{(1)} = \beta \Delta$ from the next manifold of states. The second manifold is constructed from particle-hole excitations and has a width $2 J$. If
\begin{equation}
2 J < \Delta,
\end{equation}
there is a second gap $\Delta_\rho^{(2)} = \beta (\Delta - 2 J)$ to the next manifold. This series continues, and for $N \to \infty$ we obtain gaps (assuming $\mu \ll N$)
\begin{equation}
\Delta_\rho^{(\mu)} = \beta \l \Delta - (\mu-1) 2 J \r.
\label{eq:DeltaRhoMu}
\end{equation}
In the limit of a completely flat band \cite{Kapit2010}, $J=0$, there are $N+1$ largely degenerate manifolds of states ($N$ is the number of fermions).

Now we derive the topological classification of the thermal states $\rh = e^{- \beta \H_0}/Z$. To this end we make use of the equivalence of the Chern-number and LU classifications of Bloch bands, which has been proven in Ref.~\cite{Brouder2007}. At zero temperature $T=0$, the ground state is pure and can be characterized by the many-body Chern number, $\mathcal{C}=-1$. Because the global topology of the entire spectrum is always trivial, it follows that the manifold consisting of all other states except the ground states, have total Chern number $\mathcal{C}=1$.

At finite temperature $T>0$, the spectral structure of the density matrix becomes more complex. For zero band width, $J=0$, the $\mu = 0... N$ manifolds can be characterized by their total Chern numbers, defined by integrating the Berry curvature from all bands within the manifolds. From counting one obtains 
\begin{equation}
\mathcal{C}_\mu = (- 1 +2 \mu/N) \mathcal{N}_\mu, \quad \mathcal{N}_\mu = {N \choose \mu}^2,
\end{equation}
where $\mathcal{N}_\mu$ is the number of states in the manifold labeled by $\mu$. Because a manifold with a total non-vanishing Chern number cannot be transformed to an ensemble of product states, which has a trivial Chern number, the resulting density matrix is topologically non-trivial. 

Upon variations of the band width, the spectral structure of the density matrix changes. Manifolds at large intermediate energies begin to overlap and the topological structure of the density matrix changes when the spectral gaps $\Delta_\rho^{(\mu)}$ close one after the other, see Eq.\eqref{eq:DeltaRhoMu}. Let us discuss the most extreme case when the band width $2 J > \Delta$. Now only the first and the last spectral gap $\Delta_\rho^{(1)} = \Delta_\rho^{(N)} = \beta \Delta$ remain open while rest of the spectrum is a broad continuum, see Fig.~\ref{ToyModelDMtopOrder} (c). Because the total Chern numbers of the lowest and highest states are $\mathcal{C}=-1$ and $\mathcal{C}=1$, the total Chern number of all remaining states taken together vanishes, $\mathcal{C}=0$. 

In Fig.~\ref{ToyModelDMtopOrder} (d) we show the resulting phase diagram of the toy model. At $T=0$ ($T=\infty$) the system is always in the same topologically non-trivial (trivial) equivalence class. For finite temperatures $0 < T < \infty$ different topological classes are realized, depending on the ratio of the band width to the band gap.

\subsubsection{Relation to the classical Hall effect}
As a closely related example, let us consider the (classical) Hall effect of non-interacting electrons in a magnetic field and at finite temperatures. Quantum mechanically, this situation can be understood as a thermal state $\rh_T$ of electrons occupying many different Landau levels, each of which has Chern number $\mathcal{C}=1$. Because the Landau levels are completely flat, corresponding to $J=0$ in our toy model, the density matrix has (LU) topological order at finite temperatures. The associated classical Hall current, which is directly related to the Chern number $\mathcal{C}=1$ of the Landau levels, can be understood as a direct manifestation of this density-matrix topological order.

Theoretically, the temperature can be increased further, until the electrons begin to be influenced by lattice effects in the host crystal. In this regime the same types of topological transitions as derived from our toy model are expected when the finite width of the energy bands begin to play a role. In this case, too, higher bands have negative Chern numbers \cite{HOFSTADTER1976} and the topological order of manifolds in the many-body spectrum becomes more complicated.

To study density matrix topological order of the Hall effect experimentally, we suggest to consider the Haldane model \cite{Haldane1988} at finite temperature. This model has recently been implemented using ultracold fermions in an optical lattice \cite{Jotzu2014}. By adding additional long-range tunnelings we expect that the band width can be reduced \cite{Kapit2010} and the topological transitions discussed above can be studied. Alternatively, the Hofstadter-Hubbard model can be implemented at finite temperatures and with additional interactions \cite{Sorensen2005,Hafezi2007,Aidelsburger2013,Miyake2013,Aidelsburger2014}, which we will discuss next.

\begin{figure}[b!]
\centering
\epsfig{file=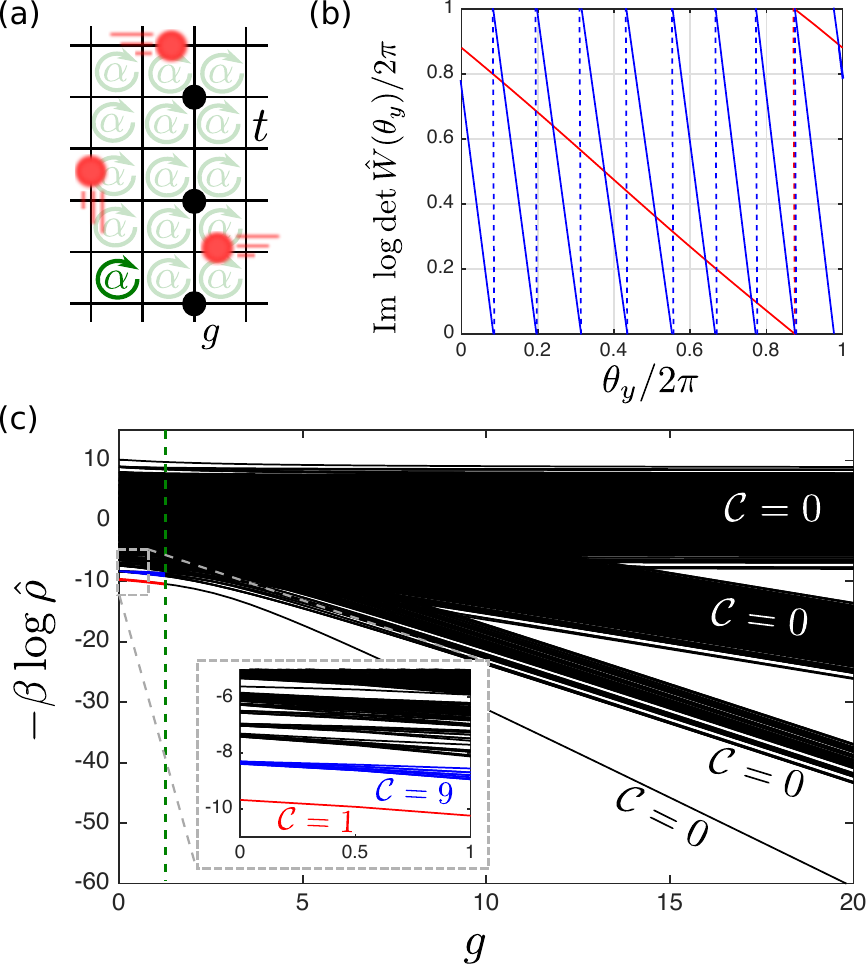, width=0.42\textwidth}
\caption{(a) We study the Hofstadter Hubbard model in a superlattice potential. Fermions with nearest neighbor interactions (strength $U$) tunnel between sites of a square lattice (hopping $t$) in a magnetic field (flux per plaquette $\alpha$). On the sites indicated by solid black dots, an attractive superlattice potential (strength $g$) is switched on, which drives a topological phase transition in the ground state for $g_c \approx 1.5 t$. We used $N=3$ particles, $\alpha=1/8$ and $U=t$ and set $t=1$. (b) The winding of the Wilson loops $W(\theta_y)$ for the first (red) and second (blue) manifold of states at $g=0$ is shown as a function of the twist angle $\theta_y$ introduced in the periodic boundary conditions. (c) The full spectrum of the thermal density matrix, $\H = - \beta \log \rh$, is shown. The spectral structure changes around $g_c \approx 1.5 t$ (green dashed line).}
\label{HofstadterHubbardFiniteT}
\end{figure}

\subsubsection{Interacting fermions in the Hofstadter-Hubbard model}
As a second example we apply the LU classification scheme to spinless thermal fermions in a Hofstadter Hubbard model \cite{HOFSTADTER1976,Sorensen2005,Hafezi2007,Aidelsburger2013,Miyake2013,Aidelsburger2014}. The Hamiltonian consists of fermion hopping between nearest neighbor sites,
\begin{equation}
 \H_t = - t \sum_{\langle i, j \rangle} e^{i \phi_{i,j}} \cd_i \c_j + \hc,
\end{equation}
with Peierls phases $e^{i \phi_{i,j}}$ giving rise to $\alpha$ units of magnetic flux per plaquette, and nearest neighbor interactions,
\begin{equation}
\H_{\rm int} = U \sum_{\langle i, j \rangle}  \cd_i \c_i \cd_j \c_j.
\end{equation}
In addition we add an attractive superlattice potential,
\begin{equation}
\H_{\rm pot} = -g \sum_j \delta_{j_x {\rm mod} 4,0} \delta_{j_y {\rm mod} 2,0}  \cd_j \c_j,
\end{equation}
as shown in Fig.~\ref{HofstadterHubbardFiniteT} (a). We consider the case when $\alpha=1/8$ at the same filling $n=1/8$.
 
For $g=0$ this model is in an integer (quantum) Hall phase and the ground state has Chern number $\mathcal{C}=1$. In Fig.~\ref{HofstadterHubbardFiniteT} (c) we show the full spectrum of the Hamiltonian. At $g=0$ we furthermore find a second manifold of $\mathcal{N}_2=9$ states, corresponding to magneto excitons. Because the first Landau level has the same Chern number as the zeroth one, we expect a total Chern number $\mathcal{C}=9$ of the second manifold. We confirmed this in Fig.~\ref{HofstadterHubbardFiniteT} (b) where we calculate the many-body Chern number of this manifold as the winding of the $U(9)$ Wilson loop \cite{Xiao2010},
\begin{equation}
\mathcal{C} = \frac{1}{2 \pi} \oint_0^{2 \pi} d\theta_y ~ \partial_{\theta_y} {\rm Im} \log \det \hat{W}(\theta_y).
\end{equation}
We use twisted periodic boundary conditions \cite{Niu1985} with twist angles $\theta_{x,y}$, and the $U(N)$ Wilson loop is defined as
\begin{equation}
\hat{W}(\theta_y) = \mathcal{P} \exp \left[ - i \int_0^{2 \pi} d \theta_x ~ \hat{\mathcal{A}}_x(\theta_x,\theta_y) \right].
\end{equation}
Here $\mathcal{P}$ defines path ordering along $\theta_x$ and $\mathcal{A}_x^{m,n}(\theta_x,\theta_y) = \bra{\psi_m(\theta_x,\theta_y)} i \partial_{\theta_{x}} \ket{\psi_n(\theta_x,\theta_y)}$ is the $U(N)$ non-Abelian Berry connection; $\ket{\psi_{m,n}(\theta_x,\theta_y)}$ for $m,n=1...N$ denote the eigenstates of the Hamiltonian from the respective manifold, at twist angles $\theta_{x,y}$.

When $g$ is increased, the ground state becomes a trivial band insulator, where all fermions are localized by the presence of the superlattice potential in the limit $g \to \infty$. By calculating its Chern number using the $U(1)$ Wilson loop, we checked that the ground state is topologically trivial with $\mathcal{C}=0$. 

The spectral structure of the density matrix changes completely when $g$ becomes large. In the limit $g \to \infty$ we expect $\mu=1...N$ manifolds above the ground state, each with $N-\mu$ fermions localized in the superlattice potential. This is confirmed by our exact numerical calculation in Fig.~\ref{HofstadterHubbardFiniteT} (c). Because of interaction effects and due to the additional superlattice potential, we obtain no additional spectral structure within these manifolds. Since they include all Landau levels in the $g \to \infty$ limit, we expect that the total Chern number of all manifolds is zero. We confirmed this by an explicit numerical calculation of the winding number of the $U(63)$ Wilson loop at $g=20 t$, corresponding to the first manifold above the ground state in this regime.

In conclusion, we have shown that the LU classification scheme allows to identify the topological order in density matrices describing correlated many-body systems. We introduced the total many-body Chern number, defined as the winding of the $U(N_\mu)$ Wilson loop corresponding to a manifold $\mu$ with $N_\mu$ states, as an efficient way to calculate topological invariants characterizing many-body density matrices.

\subsection{Relation to Diehl et al.'s scheme}
\label{sec:Diehl}
To derive the LU classification \eqref{eq:defLUdm}, we expressed the density matrix as a thermal one, with an effective Hamiltonian $\hat{h} = - \log \rh$. This analogy was exploited before by Diehl et al.~\cite{Diehl2011,Bardyn2013} who studied mixed Gaussian states of (free) fermions $\c_n$. Essentially they showed for Gaussian states that the Hamiltonian $\hat{h}$ is local, unless the system becomes critical and develops long-range correlations. Then, by distinguishing different symmetry classes of the free Hamiltonian $\hat{h} = \sum_{n,m} h_{n,m} \cd_n \c_m$ as in Ref.~\cite{Ryu2010}, they introduced topological quantum numbers which classify the bands of the single-particle Hamiltonian $h_{n,m}$. These are invariant under LU transformations \eqref{eq:LUdensityMatrices} and give rise to topologically distinct classes of mixed states. 

Formally, the key difference between Diehl et al.'s scheme and the LU classification put forward here, lies in the way how the topological properties of the Hamiltonian $\hat{h}$ are extracted, and to which Hamiltonians $\hat{h}$ the classification schemes can be applied. In Refs.\cite{Diehl2011,Bardyn2013} the Hamiltonian $\hat{h}$ needs to be quadratic in boson or fermion operators, and the single-particle Hamiltonian $h_{m,n}$ needs to be known explicitly. This allows to apply single-particle band-theory directly. In the LU scheme, in contrast, $\hat{h}$ can be an arbitrary many-body Hamiltonian, and we also do not require translational invariance of the system. To apply the LU scheme, the spectral structure of the density matrix needs to be constructed and the topology of every manifold of states is obtained, as described in Fig.~\ref{DMdefTopOrd}. Note that this procedure includes states from the entire spectrum, including high energies.

Furthermore Diehl et al.~\cite{Diehl2011,Bardyn2013} showed for Gaussian states that transitions between topologically distinct phases can take place when the system becomes critical and $\hat{h}$ is non-local ("the damping gap closes" \cite{Bardyn2013}), or when one of the gaps in the spectrum of $h_{m,n}$ (or, equivalently of $\rh$) closes ("the purity gap closes" \cite{Bardyn2013}). The same phenomenology is derived from the LU scheme: A closing of the purity gap corresponds to a closing of a gap $\Delta_\rho^{(\mu)}=0$, where the spectral structure of $\rh$ can change. When $-\log \rh$ becomes non-local, the density-matrix is no longer quasi-thermal and adiabatic deformations can induce non-local changes of the state. As a result the topological order in the sense of a pattern of long-range entanglement can change.

We conclude that the LU classification \eqref{eq:LUdensityMatrices} introduced in this paper should be understood as a generalization of the scheme developed by Diehl et al. for Gaussian states to quasi-thermal mixed states describing interacting many-body systems with strong correlations.

\section{Driven-dissipative systems}
\label{sec:OpenSystems}
So far we considered only mixed states in closed systems and with unitary perturbations. We have shown that while global topological order is robust to arbitrary local perturbations, the topology of local thermal states is only robust under weak (adiabatic) perturbations of the system. The reason was that a local subsystem can become entangled and thermalize with the surrounding parts of the closed system. Now we extend our discuss to more extreme situations, where the system is open and coupled to local baths. We are interested, in particular, in the steady state of a (driven-dissipative) open quantum system, and its dynamics when local perturbations are applied.

We consider open quantum systems coupled to Markovian baths, such that the dynamics of the density matrix can be described by a Lindblad Master equation \cite{Lindblad1976},
\begin{equation}
 \partial_t \rrh(t) = \LL(t) \rrh(t).
\end{equation}
Here $\LL(t)$ denotes the Liouville super-operator involving only local processes and acting on the vectorized density matrix $\rrh(t)$. We want to study the robustness of non-equilibrium steady states  $\rrh_0$ (for which $\uuline{\mathcal{L}_0} ~ \rrh_0 = 0$) to arbitrary local perturbations in $\LL(t)$. 

First we define more precisely the class of gapped non-equilibrium steady states (NESS) which we want to consider. We call a NESS $\rrh_0$ gapped, if it is the unique steady state of a local Liouvillian $\LL$ with a finite damping gap $\Delta_{\mathcal{L}}$; The Liouvillian is called local iff all its Lindblad generators $\hat{L}_n$ are bounded local operators. This is equivalent to the notion of gapped ground states $\ket{\psi_0}$ of local Hamiltonians in the context of closed quantum systems.

\subsection{LU classification scheme}
We will assume that $\hat{h} = - \log \rh$ is a (quasi) local operator, which -- by analogy with the result for Gaussian states \cite{Bardyn2013} -- we expect to be true when $\rh$ has no long-range correlations. In this case we can readily apply the LU classification scheme from Sec.~\ref{sec:LUtopOrderDM} to classify different topological equivalence classes of NESSs. 

Let us first discuss the effect of weak local perturbations of the Liouvillian $\LL$. When their characteristic strength $g$ is smaller than the damping gap, $g \ll \Delta_{\mathcal{L}}$, the change of the NESS is perturbative in $ \varepsilon = g / \Delta_{\mathcal{L}} \ll 1$. Such perturbative modifications of the density matrix $\rrh_0 \to \rrh_0 + \delta \rrh$ cannot close the  gaps in $- \log \rh$ defining its spectral structure when $\varepsilon$ is sufficiently small. We thus conclude that \emph{the density matrix LU topological order of a NESS of a local Liouvillian $\uuline{\mathcal{L}}$ is robust to arbitrary but weak local perturbations of $\uuline{\mathcal{L}}$.}

When the (local) perturbations of $\LL$ are strong, however, arbitrary changes of the NESS are possible on finite time scales. Consider, for example, a situation where $\LL$ is suddenly quenched to $\LL'$ with a NESS in a different phase. After a finite time determined by the damping gap of $\LL'$, i.e. $\tau \sim 1/\Delta_{\mathcal{L}'}$, the new NESS is reached. It can have completely different properties from the initial NESS we started from. Therefore \emph{LU topological order is not robust to local perturbations of arbitrary strength in an infinite system.} 

\begin{figure}[t!]
\centering
\epsfig{file=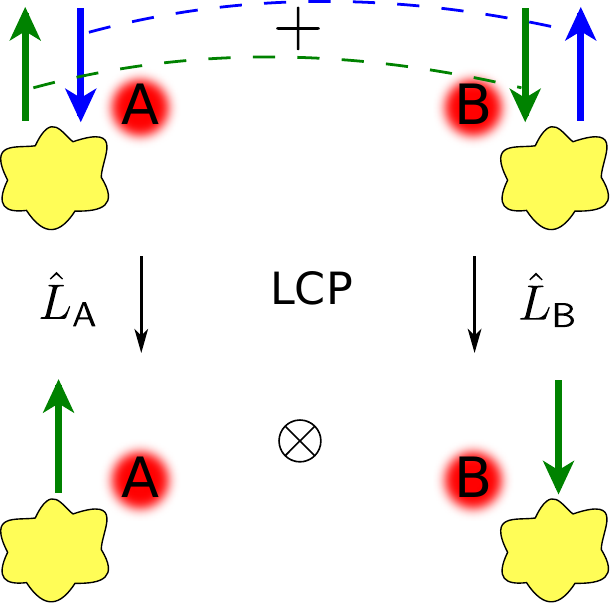, width=0.23\textwidth} $\qquad$
\caption{Topological order is not robust when couplings to local baths are considered. They can destroy the long-range entanglement in the system, which is the hallmark of topological order, in a finite time. This destruction of non-local entanglement between two parties $A$ and $B$ by the action of local Lindblad generators $\hat{L}_{A,B}$ is illustrated in the figure.}
\label{LCP_aliceBob}
\end{figure}

\subsection{Robustness of topological order -- toy model}
\label{sec:toyModelOpenSys}
Next we return to the simple toy model, which we used in the beginning (Sec.~\ref{sec:toyModel}) to illustrate the robustness of topological order of pure states in closed quantum systems. Now we consider a situation when Alice and Bob share a state $\rh$ describing a statistical mixture. We can use the eigenbasis of the density matrix and write $\rh = \sum_n \rho_n \ket{n}\bra{n}$. 

When only unitary perturbations acting separately in the systems $A$ and $B$ exist, the entanglement entropies $S_A(n)$ and $S_B(n)$ of each of the eigenstates $\ket{n}$ are conserved by LU transformations, see Eq. \eqref{eq:LUtransfAB}. Therefore $S_{A,B}(n)$ constitute good quantum numbers characterizing $\rh$, which are robust to arbitrary local unitary perturbations. This conserved non-local entanglement is the essence of the LU classification scheme for globally thermal states.

If we consider a driven-dissipative quantum system however, where Alice and Bob both have couplings to local baths, the situation changes. The dynamics of their shared quantum state $\rh(t)$ is non-unitary, described by a Lindblad Master equation if the reservoirs are Markovian. Such non-unitary time evolution no longer conserves the entanglement entropies $S_{A,B}(n)$ between the two sub-systems, even if the coupling to the baths is purely local. This is illustrated in Fig.~\ref{LCP_aliceBob}.

For concreteness, let us imagine a situation where Alice and Bob initially share the Bell state $\rh_0 = \ket{\Psi^+}\bra{\Psi^+}$ and Alice, say, measures the spin. This requires coupling to a measurement apparatus, i.e. a local reservoir, on her side. After the projective measurement, the state of the system is a statistical mixture
\begin{equation}
\rh' = \frac{1}{2} \Bigl(  \ket{\! \uparrow}_A \bra{\uparrow\!}  \otimes \ket{ \! \downarrow}_B   \bra{\downarrow \! } + \ket{ \! \downarrow}_A \bra{\downarrow \! }  \otimes \ket{ \! \uparrow}_B   \bra{\uparrow \! } \Bigr).
\end{equation}
The entanglement entropies $S(n)$ of the eigenstates of $\rh'$ both vanish, $S_A(n)=0$.

This example illustrates that topological order, which is a pattern of long-range entanglement \cite{Wen2004} of a quantum state, is not robust to local non-unitary perturbations in general. We formalize this in Appendix \ref{appendix:LCP}.

\section{Summary and Outlook}
\label{sec:outlook}
In this paper we leveraged the LU classification scheme \cite{Chen2010} to define topological order of mixed states in generic interacting quantum systems. We generalized previous results for Gaussian states of free fermions obtained by Diehl et al.~\cite{Diehl2011,Bardyn2013} and identified possible topological structures in density matrices of correlated many-body systems. According to our scheme, two density matrices are topologically equivalent if (i) they share the same spectral structure, and (ii) the ensembles of states defined by gaps in the spectrum of the density matrix are topologically equivalent; This is the case when two such ensembles can be transformed into each other by a local unitary transformation and a combination of arbitrary unitary transformations within the ensembles. We argued that this definition of topological order defines equivalence classes of density matrices which can be adiabatically transformed into each other.

Physically, the LU classification scheme distinguishes density matrices with different patterns of the long-range entanglement, which is robust to adiabatic deformations of the system. Identifying properties of mixed states which are robust to local perturbations is an important goal for quantum information applications \cite{Nayak2008}. It is relevant, in particular, for addressing the question how robust topologically protected qubits are against couplings to the environment, which are unavoidable in practice. Here we made a step in this direction by identifying such robust structures of density matrices, although the development of a topologically protected qubit in an open system is still an unsolved problem.

We have shown that the robustness of topological order in a density matrix depends crucially on the nature of the system under consideration. We pointed out that for thermal states describing global properties of a closed quantum system, density matrix topological order is robust to local perturbations of arbitrary strength in an infinite system. If only local observables are described by a reduced density matrix, however, topological order is only robust to weak local perturbations. Similarly, density matrix topological order in non-equilibrium steady states of driven dissipative open quantum systems is only robust to local perturbations which are weak compared to the damping gap.

Our work formalizes the meaning of topological order in density matrices describing generic quantum many-body systems, not restricted to quadratic Hamiltonians or Gaussian states. This paves the way for future investigations of the dynamics of topological order. In particular it allows to study local subsystems of a larger system, and investigate how topological order develops (or decays) as a function of time on different length scales. 

A key future challenge will be the direct detection of the topological order in a density matrix. As a first step, studying the reduced density matrix of a small subsystem is interesting, in particular when the global state of the system changes in time. In this case we note that one option is to use quantum state tomography to map out the entire density matrix of the small subsystem and extract its topological order afterwards.

\section*{Acknowledgements}
The author would like to thank M. Fleischhauer for initiating this research and contributing many important insights to this work. Fruitful discussions with E. Demler, M. Hafezi, E. van Nieuwenburg, S. Huber, D. Linzner, A. Turner, F. Pientka, C. Schweizer, M. Lohse and S. Diehl are also gratefully acknowledged. The author would like to thank the Gordon and Betty Moore foundation for financial support, and KITP for hospitality, where part of this work was completed. This research was supported in part by the National Science Foundation under Grant No. NSF PHY11-25915.

\appendix

\section{Local, completely positive maps}
\label{appendix:LCP}
In this appendix we discuss the analog of LU transformations for driven-dissipative open quantum systems. We will show that all gapped NESSs of local Liouvillians are equivalent according to this definition.

A natural generalization of LU transformations to open systems is to include non-unitary perturbations in the finite time evolution. We introduce local completely positive (LCP) maps as follows: \emph{A LCP map is defined by a finite-time evolution of an open quantum system according to a local Liouvillian $\tilde{\LL}(\tau)$,}
\begin{equation}
\LLCP = \mathcal{T} \exp \l - \int_0^1 d\tau  ~ \tilde{\LL}(\tau) \r.
\label{eq:defLCP}
\end{equation}

Two NESSs $\rrh_{1,2}$ can be considered equivalent under LCP transformations iff a LCP map exists such that 
\begin{equation}
\rrh_2 = \LLCP~ \rrh_1.
\end{equation}
However, because the NESS of a gapped Liouvillian is reached from any initial state after a finite time evolution, any two NESSs are equivalent under LCPs: Consider two gapped NESSs  with $\uuline{\mathcal{L}_j} ~ \rrh_j=0$. To show that $\rrh_1=\uuline{\rm LCP_1} ~ \rrh_2$, up to exponentially small corrections, the following LCP can be used,
\begin{equation}
\uuline{\rm LCP_1} = \exp ( - t_1  \uuline{\mathcal{L}_1} ) , \qquad t_1 \gg 1 / \Delta_{\mathcal{L}_1}.
\end{equation}
The finite damping gap $\Delta_{\mathcal{L}_1}>0$ allows to prepare $\rrh_1$ in a finite time. The same argument can be used to show that $\uuline{\rm LCP_2} ~ \rrh_1$. 

As a result, the analogue of LU transformations for open quantum systems, i.e. LCP maps, only define one trivial equivalence class. Thus we cannot expect to find as robust topological structures in NESSs as in pure states describing closed quantum systems.


\begin{thebibliography}{10}

\bibitem{Ginzburg1950}
V.~L. Ginzburg and L.~D. Landau.
\newblock To the theory of superconductivity.
\newblock {\em Zh. Eksp. Teor. Fiz.}, 20:1064, 1950.

\bibitem{Vonklitzing1980}
K.~Von~Klitzing, G.~Dorda, and M.~Pepper.
\newblock New method for high-accuracy determination of the fine-structure
  constant based on quantized hall resistance.
\newblock {\em Physical Review Letters}, 45(6):494--497, 1980.

\bibitem{Laughlin1981}
R.~B. Laughlin.
\newblock Quantized hall conductivity in 2 dimensions.
\newblock {\em Physical Review B}, 23(10):5632--5633, 1981.

\bibitem{Thouless1982}
D.~J. Thouless, M.~Kohmoto, M.~P. Nightingale, and M.~denNijs.
\newblock Quantized hall conductance in a two-dimensional periodic potential.
\newblock {\em Physical Review Letters}, 49(6):405--408, 1982.

\bibitem{KOHMOTO1985}
M.~Kohmoto.
\newblock Topological invariant and the quantization of the hall conductance.
\newblock {\em Annals of Physics}, 160(2):343--354, 1985.

\bibitem{Wen2004}
Xiao-Gang Wen.
\newblock {\em Quantum Field Theory of Many-body Systems}.
\newblock Oxford University Press, 2004.

\bibitem{Chen2010}
Xie Chen, Zheng-Cheng Gu, and Xiao-Gang Wen.
\newblock Local unitary transformation, long-range quantum entanglement, wave
  function renormalization, and topological order.
\newblock {\em Physical Review B}, 82(15):155138, October 2010.

\bibitem{Wen1995}
X.~G. Wen.
\newblock Topological orders and edge excitations in fractional quantum hall
  states.
\newblock {\em Advances In Physics}, 44(5):405--473, 1995.

\bibitem{Laughlin1983}
R.~B. Laughlin.
\newblock Anomalous quantum hall-effect - an incompressible quantum fluid with
  fractionally charged excitations.
\newblock {\em Physical Review Letters}, 50(18):1395--1398, 1983.

\bibitem{Kitaev2006}
A.~Kitaev and J.~Preskill.
\newblock Topological entanglement entropy.
\newblock {\em Physical Review Letters}, 96(11):110404, March 2006.

\bibitem{Ryu2010}
Shinsei Ryu, Andreas~P. Schnyder, Akira Furusaki, and Andreas Ludwig.
\newblock Topological insulators and superconductors: tenfold way and
  dimensional hierarchy.
\newblock {\em New Journal of Physics}, 12:065010, 2010.

\bibitem{Diehl2011}
Sebastian Diehl, Enrique Rico, Mikhail~A. Baranov, and Peter Zoller.
\newblock Topology by dissipation in atomic quantum wires.
\newblock {\em Nature Physics}, 7(12):971--977, 2011.

\bibitem{Bardyn2013}
C.-E. Bardyn, M.~A. Baranov, C.~V. Kraus, E.~Rico, A.~Imamoglu, P.~Zoller, and
  S.~Diehl.
\newblock Topology by dissipation.
\newblock {\em New Journal of Physics}, 15:085001, 2013.

\bibitem{Uhlmann1986}
Armin Uhlmann.
\newblock Parallel transport and "quantum holonomy" along density
  operators.
\newblock {\em Reports on Mathematical Physics}, 24:229, 1986.

\bibitem{Avron2011}
J~E Avron, M~Fraas, G~M Graf, and O~Kenneth.
\newblock Quantum response of dephasing open systems.
\newblock {\em New Journal of Physics}, 13(5):053042, 2011.

\bibitem{Avron2012}
J.E. Avron, M.~Fraas, and G.M. Graf.
\newblock Adiabatic response for lindblad dynamics.
\newblock 148(5):800--823--, 2012.

\bibitem{Rivas2013a}
A.~Rivas, O.~Viyuela, and M.~A. Martin-Delgado.
\newblock Density-matrix chern insulators: Finite-temperature generalization of
  topological insulators.
\newblock {\em Physical Review B}, 88(15):155141, October 2013.

\bibitem{Huang2014}
Zhoushen Huang and Daniel~P. Arovas.
\newblock Topological indices for open and thermal systems via uhlmann's phase.
\newblock {\em Phys. Rev. Lett.}, 113:076407, Aug 2014.

\bibitem{Nieuwenburg2014}
Evert P.~L. van Nieuwenburg and Sebastian~D. Huber.
\newblock Classification of mixed-state topology in one dimension.
\newblock {\em Phys. Rev. B}, 90:075141, Aug 2014.

\bibitem{Viyuela2014}
O.~Viyuela, A.~Rivas, and M.~A. Martin-Delgado.
\newblock Two-dimensional density-matrix topological fermionic phases:
  Topological uhlmann numbers.
\newblock {\em Phys. Rev. Lett.}, 113:076408, Aug 2014.

\bibitem{Budich2015}
Jan~Carl Budich, Peter Zoller, and Sebastian Diehl.
\newblock Dissipative preparation of chern insulators.
\newblock {\em Phys. Rev. A}, 91:042117, Apr 2015.

\bibitem{Budich2015a}
Jan~Carl Budich and Sebastian Diehl.
\newblock Topology of density matrices.
\newblock {\em Phys. Rev. B}, 91:165140, Apr 2015.

\bibitem{Albert2016}
Victor~V. Albert, Barry Bradlyn, Martin Fraas, and Liang Jiang.
\newblock Geometry and response of lindbladians.
\newblock {\em arXiv:1512.08079v3}.

\bibitem{Linzner2016}
D.~Linzner, L. Wawer, F.~Grusdt, and M.~Fleischhauer.
\newblock Thouless pumping and reservoir-induced topological order in
  interacting open spin chains.
\newblock {\em arXiv:1605.00756}.

\bibitem{Vidal2004}
Guifr\'e Vidal.
\newblock Efficient simulation of one-dimensional quantum many-body systems.
\newblock {\em Phys. Rev. Lett.}, 93:040502, Jul 2004.

\bibitem{Verstraete2004}
F.~Verstraete, J.~J. Garcia-Ripoll, and J.~I. Cirac.
\newblock Matrix product density operators: Simulation of finite-temperature
  and dissipative systems.
\newblock {\em Phys. Rev. Lett.}, 93:207204, Nov 2004.

\bibitem{Eisert2010RMP}
J.~Eisert, M.~Cramer, and M.~B. Plenio.
\newblock Colloquium: Area laws for the entanglement entropy.
\newblock {\em Rev. Mod. Phys.}, 82:277--306, Feb 2010.

\bibitem{Hastings2011}
Matthew~B. Hastings.
\newblock Topological order at nonzero temperature.
\newblock {\em Phys. Rev. Lett.}, 107:210501, Nov 2011.

\bibitem{Berry1984}
M.~V. Berry.
\newblock Quantal phase-factors accompanying adiabatic changes.
\newblock {\em Proceedings of the Royal Society of London Series A-mathematical
  Physical and Engineering Sciences}, 392(1802):45--57, 1984.

\bibitem{Zak1989}
J.~Zak.
\newblock Berrys phase for energy-bands in solids.
\newblock {\em Physical Review Letters}, 62(23):2747--2750, June 1989.

\bibitem{Xiao2010}
Di~Xiao, Ming-Che Chang, and Qian Niu.
\newblock Berry phase effects on electronic properties.
\newblock {\em Reviews of Modern Physics}, 82(3):1959--2007, July 2010.

\bibitem{Note1}
If there is no gap, a different gauge choice can be used to obtain a different
  Chern number, making the latter ill-defined.

\bibitem{Hastings2007}
M~B Hastings.
\newblock An area law for one-dimensional quantum systems.
\newblock {\em Journal of Statistical Mechanics: Theory and Experiment},
  2007(08):P08024--, 2007.

\bibitem{Brouder2007}
Christian Brouder, Gianluca Panati, Matteo Calandra, Christophe Mourougane, and
  Nicola Marzari.
\newblock Exponential localization of wannier functions in insulators.
\newblock {\em Phys. Rev. Lett.}, 98:046402, Jan 2007.

\bibitem{Abanin2012Ent}
Dmitry~A. Abanin and Eugene Demler.
\newblock Measuring entanglement entropy of a generic many-body system with a
  quantum switch.
\newblock {\em Phys. Rev. Lett.}, 109:020504, Jul 2012.

\bibitem{Daley2012}
A.~J. Daley, H.~Pichler, J.~Schachenmayer, and P.~Zoller.
\newblock Measuring entanglement growth in quench dynamics of bosons in an
  optical lattice.
\newblock {\em Phys. Rev. Lett.}, 109:020505, Jul 2012.

\bibitem{Islam2015}
Rajibul Islam, Ruichao Ma, Philipp~M. Preiss, M.~Eric~Tai, Alexander Lukin,
  Matthew Rispoli, and Markus Greiner.
\newblock Measuring entanglement entropy in a quantum many-body system.
\newblock {\em Nature}, 528(7580):77--83, December 2015.

\bibitem{Pichler2016}
Hannes Pichler, Guanyu Zhu, Alireza Seif, Peter Zoller, and Mohammad Hafezi.
\newblock A measurement protocol for the entanglement spectrum of cold atoms.
\newblock {\em arXiv:1605.08624}.

\bibitem{Einstein1935}
A.~Einstein, B.~Podolsky, and N.~Rosen.
\newblock Can quantum-mechanical description of physical reality be considered
  complete?
\newblock {\em Phys. Rev.}, 47:777--780, May 1935.

\bibitem{Bravyi2006}
S.~Bravyi, M.~B. Hastings, and F.~Verstraete.
\newblock Lieb-robinson bounds and the generation of correlations and
  topological quantum order.
\newblock {\em Phys. Rev. Lett.}, 97:050401, Jul 2006.

\bibitem{Rigol2012}
Marcos Rigol and Mark Srednicki.
\newblock Alternatives to eigenstate thermalization.
\newblock {\em Phys. Rev. Lett.}, 108:110601, Mar 2012.

\bibitem{HOFSTADTER1976}
D.R. Hofstadter.
\newblock Energy-levels and wave-functions of bloch electrons in rational and
  irrational magnetic-fields.
\newblock {\em Physical Review B}, 14(6):2239--2249, 1976.

\bibitem{Haldane1988}
F.~D.~M. Haldane.
\newblock Model for a quantum hall-effect without landau-levels -
  condensed-matter realization of the parity anomaly.
\newblock {\em Physical Review Letters}, 61(18):2015--2018, 1988.

\bibitem{Kapit2010}
Eliot Kapit and Erich Mueller.
\newblock Exact parent hamiltonian for the quantum hall states in a lattice.
\newblock {\em Physical Review Letters}, 105(21):215303, November 2010.

\bibitem{Jotzu2014}
Gregor Jotzu, Michael Messer, Remi Desbuquois, Martin Lebrat, Thomas Uehlinger,
  Daniel Greif, and Tilman Esslinger.
\newblock Experimental realization of the topological haldane model with
  ultracold fermions.
\newblock {\em Nature}, 515(7526):237--240, November 2014.

\bibitem{Sorensen2005}
A.~S. Sorensen, E.~Demler, and M.~D. Lukin.
\newblock Fractional quantum hall states of atoms in optical lattices.
\newblock {\em Physical Review Letters}, 94(8):086803, 2005.

\bibitem{Hafezi2007}
M.~Hafezi, A.~S. Sorensen, E.~Demler, and M.~D. Lukin.
\newblock Fractional quantum hall effect in optical lattices.
\newblock {\em Physical Review A}, 76(2):023613, 2007.

\bibitem{Aidelsburger2013}
M.~Aidelsburger, M.~Atala, M.~Lohse, J.~T. Barreiro, B.~Paredes, and I.~Bloch.
\newblock Realization of the hofstadter hamiltonian with ultracold atoms in
  optical lattices.
\newblock {\em Physical Review Letters}, 111:185301, 2013.

\bibitem{Miyake2013}
H.~Miyake, G.~A. Siviloglou, C.~J. Kennedy, W.~C. Burton, and W.~Ketterle.
\newblock Realizing the harper hamiltonian with laser-assisted tunneling in
  optical lattices.
\newblock {\em Physical Review Letters}, 111:185302, 2013.

\bibitem{Aidelsburger2014}
M.~Aidelsburger, M.~Lohse, C.~Schweizer, M.~Atala, J.~T. Barreiro,
  S.~Nascimbene, N.~R. Cooper, I.~Bloch, and N.~Goldman.
\newblock Measuring the chern number of hofstadter bands with ultracold bosonic
  atoms.
\newblock {\em Nat Phys}, 11(2):162--166, February 2015.

\bibitem{Niu1985}
Q.~Niu, D.~J. Thouless, and Y.~S. Wu.
\newblock Quantized hall conductance as a topological invariant.
\newblock {\em Physical Review B}, 31(6):3372--3377, 1985.

\bibitem{Lindblad1976}
G.~Lindblad.
\newblock On the generators of quantum dynamical semigroups.
\newblock {\em Communications in Mathematical Physics}, 48(2):119--130, 1976.

\bibitem{Nayak2008}
Chetan Nayak, Steven~H. Simon, Ady Stern, Michael Freedman, and Sankar
  Das~Sarma.
\newblock Non-abelian anyons and topological quantum computation.
\newblock {\em Reviews of Modern Physics}, 80(3):1083--1159, 2008.

\end{thebibliography}

\end{document}